\documentstyle[aps,preprint]{revtex}
\input epsf

\title{Cavitation induced by explosion in a model of ideal fluid}

\author{Christophe Josserand\\
\small\it The James Franck Institute,\\
The University of Chicago,\\
5640 South Ellis Avenue,\\
Chicago, Illinois 60637, USA
}

\author{\parbox{430pt}{\vglue 0.3 cm \small We discuss the problem of
an explosion in the cubic-quintic superfluid model, in relation to some
experimental observations. We show numerically that an explosion in
such a model might induce a cavitation bubble for large enough energy.
This gives a consistent view for rebound bubbles in superfluid and we
indentify the loss of energy between the successive rebounds as
radiated waves. We
compute self-similar solution of the explosion for the early stage,
when no bubbles have been nucleated. The solution also gives the 
wave number of the excitations emitted through the shock wave.}}

\begin{document}

\maketitle
\tableofcontents

\section{Introduction.}

Cavitation is a physical process involving such aspects as erosion,
bubble formation, sonoluminescence\cite{sono1} and first order phase
transitions. The motivation for this work comes from experiments on
superfluid Helium\cite{bali1,bali2} and we will generally speak
 in this context, although the model might be applied to others fluids,
 and to non linear optics\cite{newell}. The experiments of
 \cite{bali1,bali2} study the cavitation process in superfluid Helium
(He$^4$). There, a semi-spherical convergent sound wave is produced in
the liquid. At the center of the set-up, the superfluid alternates
between being compressed and being under tension ({\it i.e.} at
negative pressure). During the tensile strength period, a bubble can be
nucleated by thermal activation.  This bubble acquires a kinetic energy
through the negative pressure region so that it grows inside the liquid
bulk until it reaches a maximal radius (a few hundred $\mu m$ can be
obtained ) determined by its kinetic energy and the mean liquid
pressure.  Then, the bubble collapses under this positive pressure.
After its collapse, a secondary bubble, called the rebound bubble is
observed. The collapse of bubbles is a catastrophic process that has
been widely studied, particularly because of its ramification in
industrials application (for a review, see \cite{brennen}).
Sonoluminescence arises also during the collapse of the
bubble\cite{lauterborn}. The collapse is generally followed by a shock
wave that is often a cause of important damages.  The goal of this
paper is to show that in the particular case of superfluid, the rebound
bubble can actually be nucleated by this shock wave. It has been
noticed already that detonation in water can nucleate cavitation bubble
through the tension shock wave that follow an explosion (see
\cite{lauter}).

We will first introduce a model recently used for phase transition in
systems like superfluids\cite{bulle,goutte}.  Then we will numerically
study the problem of explosion in this model as the collapse of the
bubble might indeed be interpreted as an explosion. The bubble that
will form beside the shock wave might be understood as the rebound
bubble found in the experiments. We show also that the explosion
process exhibits two distinct regimes: for small time, it follows a
pure explosion behavior whereas an interaction between sound wave
propagation and interface dynamics dictates the larger times. We
finally examine the large energy limit, where we can expand the explosion
in terms of a self-similar solution for small time.

\section{The Gross-Pitaevski\u{\i} equation.}
The Gross-Pitaevski\u{\i} (GP) equation has been often used as a model of 
superfluid\cite{pit}.
 It describes the time evolution of a complex function $\psi({\bf x};t)$,
 called the condensate wave function; it reads: 
\begin{equation}
 \imath \hbar \partial_t \psi({\bf r},t)=-\frac{\hbar^2}{2m} \psi({\bf r},t)
+g|\psi({\bf r},t)|^2\psi({\bf r},t) 
\label{GP}
\end{equation}
where $\hbar$ is the Planck constant, $m$ the mass of the particles (for 
superfluid, this is $m_{He^4}$) and $g$ is the strength of the potential.\\
One can write:
$$ \psi=\sqrt{\rho}e^{i\phi} $$
and in analogy with quantum mechanics terminology, one can identify 
$\rho$ as the particle's density and $\frac{\hbar}{m}\phi$ as the velocity potential:
$${\bf v}({\bf r},t)= \frac{\hbar}{m}{\bf \nabla}\phi({\bf r},t) $$
It follows that the equation is conservative: the total number of particles $N$ is
conserved by the dynamics. 
$$ N=\int d{\bf r}|\psi|^2 $$

The dynamics is also hamiltonian which means that one can define an
energy $H$ such that:

$$ i\hbar \partial_t\psi=\frac{\delta H}{\delta \psi^*} $$

where $\psi^*$ is the complex conjugate of $\psi$ and with:

$$ H=\int d{\bf r} \left( \frac{\hbar^2}{2m}|{\bf \nabla}\psi|^2+\frac{g}{2}
|\psi|^4 \right) $$
The halmitonian structure implies that the dynamics is reversible.

A set of equations of  hydrodynamic form (for $\rho$ and $\phi$) can be
deduced from the GP equation:

\begin{eqnarray}
\partial_t\rho +{\bf \nabla}\cdot(\rho{\bf v})&=&0 \nonumber \\ 
-\partial_t\left(\frac{\hbar \phi}{m}\right)&=&-\frac{\hbar^2}{2m^2}\cdot
\frac{\Delta \sqrt{\rho}}{\sqrt{\rho}}+\frac{1}{2}v^2+\frac{g}{m}\rho 
\nonumber \\
\nonumber
\end{eqnarray}

The first equation is the mass conservation and the second one can be
viewed as an equivalent to the Bernoulli equation for fluids. Indeed
the pressure is divided into two terms: the first one, gives a static
pressure: $P=\frac{g}{2m} \rho^2$ whereas $\frac{\Delta
\sqrt{\rho}}{\sqrt{\rho}}$ is called the quantum pressure term, because
it annihilates when $\hbar \rightarrow 0$.  This term is a reminiscence
that the dynamic is deduced from a Schr\"{o}dinger equation. Without
this term, the equation would read exactly as the Euler equation for
perfect fluid with a given state equation for $P(\rho)$.  In
fact, the quantum pressure changes this Euler dynamics, (while conserving the
total energy) whereas for real fluid, it is the viscosity that
stabilizes the flow, with a dissipative dynamics.

Actually, no damping terms (like viscous term) are present in this
dynamics; this is necessary for a consistent model of superfluid; also,
 this equation admits only liquid-like solutions ($\psi=\sqrt{\rho_0}
e^{-i\frac{g\rho_0}{\hbar}t}$, where $\rho_0$ is the liquid density,
constant); the perturbations around such solution:
$$\psi=\left(\sqrt{\rho_0}+\delta\psi e^{i ( \omega \cdot t-{\bf
p}\cdot {\bf r})/\hbar}\right)e^{-i\frac{g\rho_0}{\hbar}t} $$
 respects the dispersion equation:  $$ \omega^2= \frac{g \cdot
\rho_0}{m} p^2+\frac{p^4}{4m^2} $$ $\omega$ being the energy of the
so-called quasiparticles and ${\bf p}$ their momentum ($p=\hbar \cdot
k$, $k$ wavenumber).  For low wave numbers, the quasiparticles are
phonons, {\it i.e.} sound waves with sound velocity $c_s=\sqrt{\frac{g
\cdot \rho_0}{m}}$. The phonon spectrum for low $p$ has been pointed
out to be a crucial property needed for modeling superfluids, according
to Landau theory\cite{lan41}.  For large wave numbers, the spectrum
corresponds to the one for free particles (the kinetic term is
dominant). Then, unfortunately the model does not describe the roton
part of the superfluid spectrum, although a non local potential in
equation (\ref{GP}) would allow for such a spectrum\cite{roton}.  The
crossover between the two regimes occurs for a typical length, called
the coherence length $\xi_0$:  
$$ \xi_0=\frac{\hbar}{\sqrt{mg\rho_0}}$$

Finally, such a system contains by construction another important
feature of superfluidity, the so-called quantum vortices. They are in
fact topological defects associated to the complex order parameter
$\psi$ and reflect that its phase might be multivalued (modulo
$2\pi$).  As the phase of $\psi$ is related to the velocity potential,
the circulation of the vortices is a multiple of $h/m$, as predicted by
Onsager for superfluid \cite{ons}.

Briefly, all these properties have made the Gross-Pitaevski\u{\i}
equation a reasonable model for a superfluid at $T=0$ Kelvin, and it is
often used due to its balance between simplicity and sufficient
physical ingredients.Therefore it has been particularly used for
numerical studies. This equation might also be simply considered as a
fluid dynamics model, satisfying an Euler equation but with an
additional term arising from the quantum pressure. This term becomes
relevant for dynamics on length scales smaller than the coherence
length $\xi_0$, and in particular it stabilizes the vortex core.  Such
a fluid model presents an alternative view of fluid dynamics, in that
divergences (shocks...)  are eliminated by dispersion rather than by
dissipation (as the viscosity does for real fluids). Aside from this
theoretical point of view, such approach might give interesting
prospects for real fluids.

In the context of this paper, it is relevant to introduce a 
dimensionless form of equation (\ref{GP}). It is the so-called non linear 
Schr\"{o}dinger equation (NLS), obtained by trivial rescaling of the space, the
time and the mass; it reads:
\begin{equation}
\imath \partial_t \psi = -\frac{1}{2} \Delta \psi +|\psi|^2 \psi
\label{nls}
\end{equation}

Therefore, if $\rho_0$ is the mean density of particles, the sound 
velocity is worth $\sqrt{\rho_0}$, the coherence length $\xi_0=1/\sqrt{\rho_0}$,
the typical time scale being $1$.

\section{The model}

The equation above describes, in fact, the dynamics of a monophase
fluid: precisely the model allows only one thermodynamically stable
phase, called liquid phase.  Therefore the model is not relevant for
any problems involving liquid-gas transition or any first order phase
transition. Particularly, all the interaction that can occur in high
speed flow between vorticity and cavitation are lost in such an
approach. It has been shown that a small change in the NLS equation can
give a consistent first order liquid-gas transition without changing
the other properties of NLS; it has then be used for studying standard
problems like flow around an obstacle in two space dimensions and
coarsening processes\cite{bulle,goutte}. This subcritical non linear
Schr\"{o}dinger (SNLS) equation reads in a dimensionless form:

\begin{equation}
\imath \partial_t \psi = -\frac{1}{2} \Delta \psi +(2 \rho_c-\rho_0)\rho_0 \psi
-2\rho_c |\psi|^2 \psi +|\psi|^4 \psi
\label{snls}
\end{equation} 
$\rho_c$ and $\rho_0$ are respectively the critical and the mean densities as  
 will be explained below. \\
The equivalent set of equation for (\ref{snls}) reads:
\begin{eqnarray}
\partial_t \rho & = & -{\bf \nabla} \cdot (\rho {\bf v}) \label{conts}\\
\partial_t \phi &= & \frac{1}{2 \rho^{1/2}} \Delta (\rho^{1/2})-\frac{1}{2}
({\bf \nabla} \phi)^2-\rho^2+2\rho_c \rho -\rho_0(2 \rho_c-\rho_0) 
\label{berns}
\end{eqnarray}
They have the same structure than the one deduced for the Gross-Pitaevski\u{\i}
equation.

The Bernoulli equation allows us to define the static pressure $P$
(forgetting the quantum pressure contribution) and therefore a sound
velocity $c_s$ might also be derived:

$$ P=\frac{2}{3}\rho^3-\rho_c \rho^2; \quad \quad c^2=2\rho(\rho-\rho_c) $$

\begin{figure}
\centerline{ \epsfxsize=16truecm \epsfbox{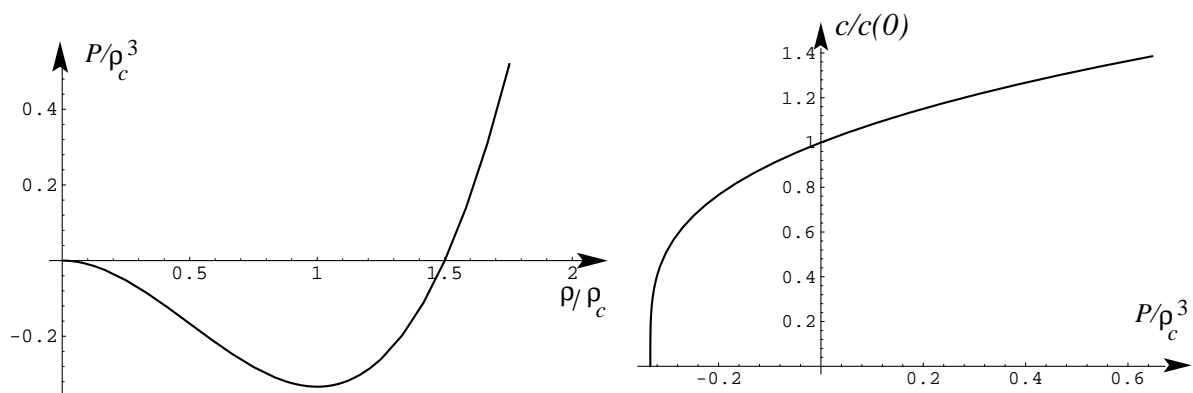}}
\caption{\protect\small Respectively, the pressure as function of the density 
and the sound velocity as function of the pressure for the model SNLS. 
These graphs show a consistent liquid-gas transition, where the spinodal 
decomposition point is located at $\rho=\rho_c$.
\label{preson}}
\end{figure}

They are shown on figure (\ref{preson}). It appears clearly that
$\rho_c$ corresponds to the spinodal decomposition density, where the
sounds velocity vanishes. As no temperature exists in the model, the
pressure dependence in $\rho$ plays the role of a state equation for
the fluid. Particularly, with $\rho_0$ being the liquid density, one can
investigate the pressure difference between the liquid phase and the
gas phase (which is at zero pressure). For $\rho_0<\frac{3}{2}\rho_c$ the
liquid pressure is smaller than the gas pressure, which means that the
liquid phase is metastable relative to the gas one; we have the opposite
situation for $\rho_0<\frac{3}{2}\rho_c$. $\rho=\frac{3}{2}\rho_c$ is
the density for which the liquid and the gas pressures are equal.
Therefore, the density of the liquid when gas and liquid coexist has to
be $\rho=\frac{3}{2}\rho_c$.

In one dimension, exact solutions of equation (\ref{snls}) are known
for a given number of particles (see \cite{goutte}). Consequently, the
gas-liquid interface can be exactly computed. It connects a region of
zero density to a region of $\frac{3}{2}\rho_c$ density. The energy of
such a solution gives the surface tension $\alpha$ ({\it i.e.} the
energy of the liquid/gas interface) of (\ref{snls}):

$$ \alpha= \frac{9 \rho_c^2}{16 \sqrt{6}}  $$ 

We have available now a mean field model of a first order transition.
The fluid obeys an Euler-like equation of motion {\it via} a complex
order parameter $\psi$. Even though the model is not entirely
physically realistic, it has been used for quantum flows, and we
believe that it has some important qualitative features. Particularly,
as a mean field model, the liquid-gas interface is automatically solved
by equation (\ref{snls}) without using special treatment.

\section{The explosion}  
\label{sec2}

Using the model introduced above, we will now focus on the 
particular problem of strong explosion. Such questions have been
investigated long ago for gases \cite{landau} and interesting
self similar dynamics have been pointed out. As proposed in the introduction, 
we will consider that the collapse of the initial bubble in the experiment 
leads to the formation of an explosion in the bulk. In SNLS, such an explosion 
corresponds to a peak of energy (or equivalently of density) centered at the origin
 of the collapse.
Numerically, the initial conditions for the explosion with a peak of density 
centered in $r=0$ will be taken as:

$$ \psi(r,t=0)=\sqrt{\rho_0}+\frac{\delta}{cosh(\frac{r}{\sigma})} $$

Here, $\rho_0$ is the liquid density, which is chosen such that the liquid is the
stable equilibrium phase (the gas being the metastable one). For now on
 we will have $\rho_0=1$ and $\rho_c=0.6$ (therefore all the physical
quantities are of order one). $\delta$ is the amplitude of the
excess pressure due to the explosion and $\sigma$ is its width. For a
strong explosion, where $\delta^2 \gg 1$, the energy $E_0$ of the
explosion is given by:

$$ E_0=4 \pi \int_0^{\infty} \left( \left(\frac{\delta}{\sigma} \right)^2 \frac{
{\rm sinh}^2(r/\sigma)}{{\rm cosh}^4(r/\sigma)}+\frac{1}{3} \cdot
\frac{\delta^6}{{\rm cosh}^6(r/\sigma)} \right)r^2 dr =
\frac{2}{3}(\delta^2\sigma)+\frac{16}{15} (\delta^2\sigma)^3 $$

Physically, $\sigma$ has to be on the order of few coherence lengths (we 
will show mainly results with $\sigma=1$ although we did try a large range of 
values). Also it 
appears that the process we will describe below is robust and does not
depend strongly on the specific initial conditions.

The numerical simulations have been performed with a finite difference
Crank-Nicholson scheme\cite{crank}, that preserves the number of
particles exactly at the first order in time. The code has been
employed for $1$, $2$ and $3$ space dimensions with cylindrical and
spherical symmetry for $2$ and $3$ dimensions respectively. We did
in fact neglect the Rayleigh-Taylor (RT) instability during the
simulations. The RT instability appears generally only during the
collapse and is enhanced by the proximity of a boundary (see
\cite{brennen}). However, whether the collapse is symmetric or not, it will still
give rise to an explosion (perhaps weaker in the asymmetric case).

One of the main numerical limitation comes from the fact that the sound
velocity is proportional to the local density for large densities $\rho
\gg \rho_0$; therefore, for investigating strong explosions, one needs to
deal with large sound velocity in the initial density peak.

Figure (\ref{evolu}) shows the evolution of the density profile in
spherical geometry for $\delta=3$ and $\sigma=1$.  As expected, a gas
bubble is nucleated backwards the shock. Also, a train of spherical
waves is emitted during the process. The bubble grows until a
maximal radius is reached and then collapse occurs. In figure d), the bubble has just
collapsed, giving rise to a secondary explosion, much smaller than the
initial one, because part of the energy has been transformed in excitations
waves. This latest stage gives a convincing proof that the collapse of a
bubble in such a model might be investigated as an explosion. The
bigger the initial explosion is, the bigger the maximal radius and the
secondary explosion are. Also, if the initial energy is lowered, no
bubbles are nucleated below a certain critical energy.  Formally, the
secondary explosion can nucleate an new bubble and so on as long as it
has enough energy, although we did not investigate large enough initial
explosions to see at least a secondary bubble.

Whether or not a bubble is nucleated, the explosion in SNLS exhibits a
general picture: the explosion expands until a maximum radius and then
collapse in a secondary explosion, which expands again until its
maximum radius (smaller than the former one) and so on. It describes
therefore an oscillating process where the energy of the explosion
decreases each cycle, due to the emission of waves during the
explosion.

These numerical simulations have been also computed in bidimensional
geometry (cylindrical explosion) and in a one dimensional system
(planar explosion). It appears that the cavitation process occurs for
cylindrical waves but not for planar explosion. An explanation from
classical results on irrotational flows can be advanced\cite{landau1}:
in cylindrical and spherical geometries, sound waves always contain
both a compressive and a tensile phase, whereas a planar compression
wave might occur without being accompanied by tension. Then, the
explosion in two and three spatial dimensions give rise to a negative
pressure region where the cavitation can take place if the tension is
big enough; nothing comparable happens in one dimension.

\section{Decomposition of the process}

The feature of the explosion can be captured by the evolution of the
point where the density reaches its greatest value. The density
$\rho_p$ and the position $R_p$ of this point (called P) will be, in
particular, investigated. We will indeed focus on large initial
density peak, ({\it i.e.} $\delta^2 \gg 1$); it corresponds typically to
an initial pressure peak of more than one hundred bars in superfluid
helium. In the experiment described above,
 such pressure cannot be obtained by convergent sound waves but only
 when the primitive bubble collapses\cite{comm}.

\begin{figure}[h]
\centerline{ \epsfxsize=16truecm \epsfbox{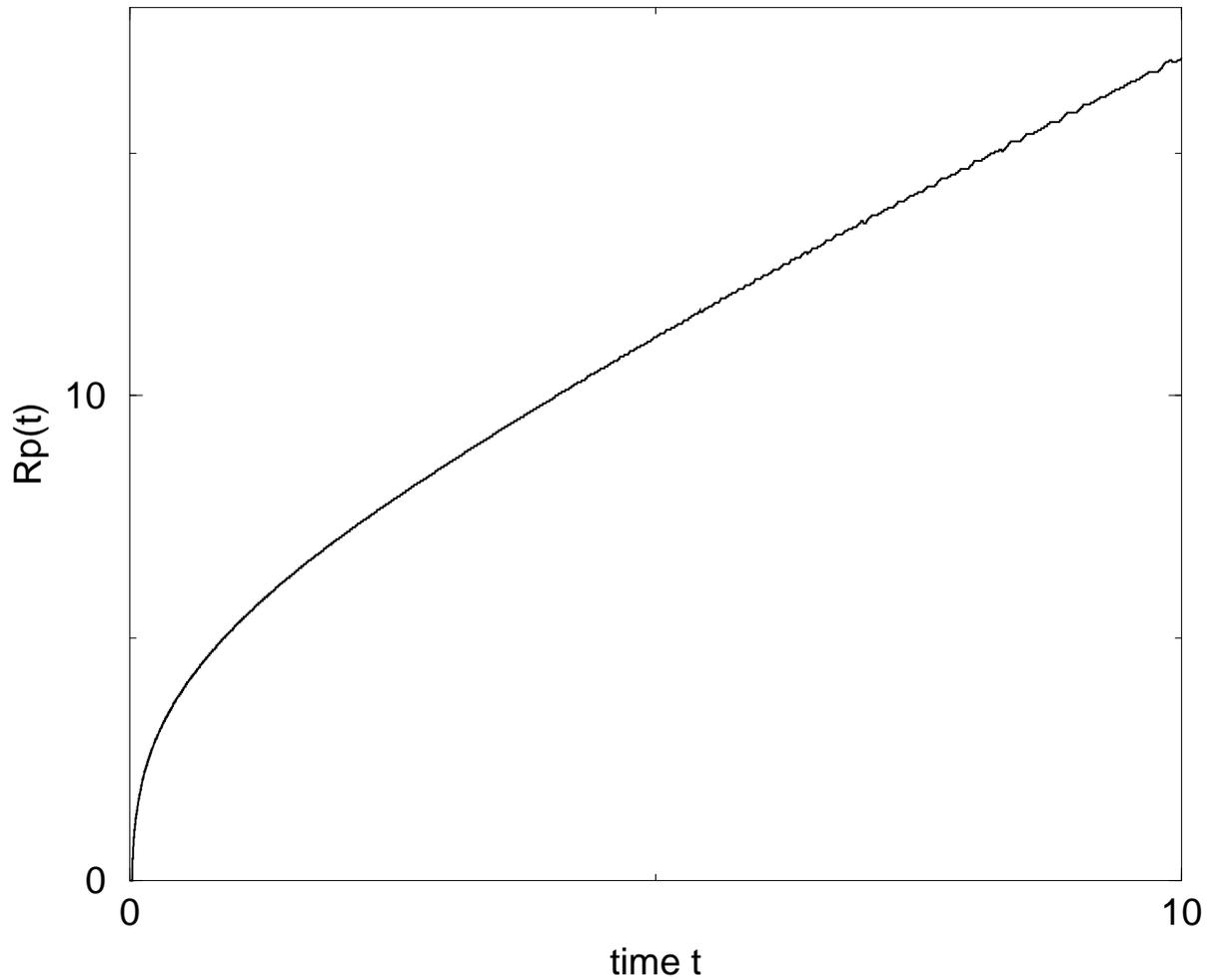} }
\caption{\protect\small Position, as a function of time, of the maximum
of the density. It exhibits two typical behavior, one for small time,
when the wave is supersonic, the other one later, where the shock wave
has transformed in a sound wave (the position evolves then linearly in
time). For the small time, the position of the shock wave is consistent
with a law $R_p \propto t^{1/2}$.
\label{maxi}}
\end{figure}

\begin{figure}[h]
\centerline{ \epsfxsize=16truecm \epsfbox{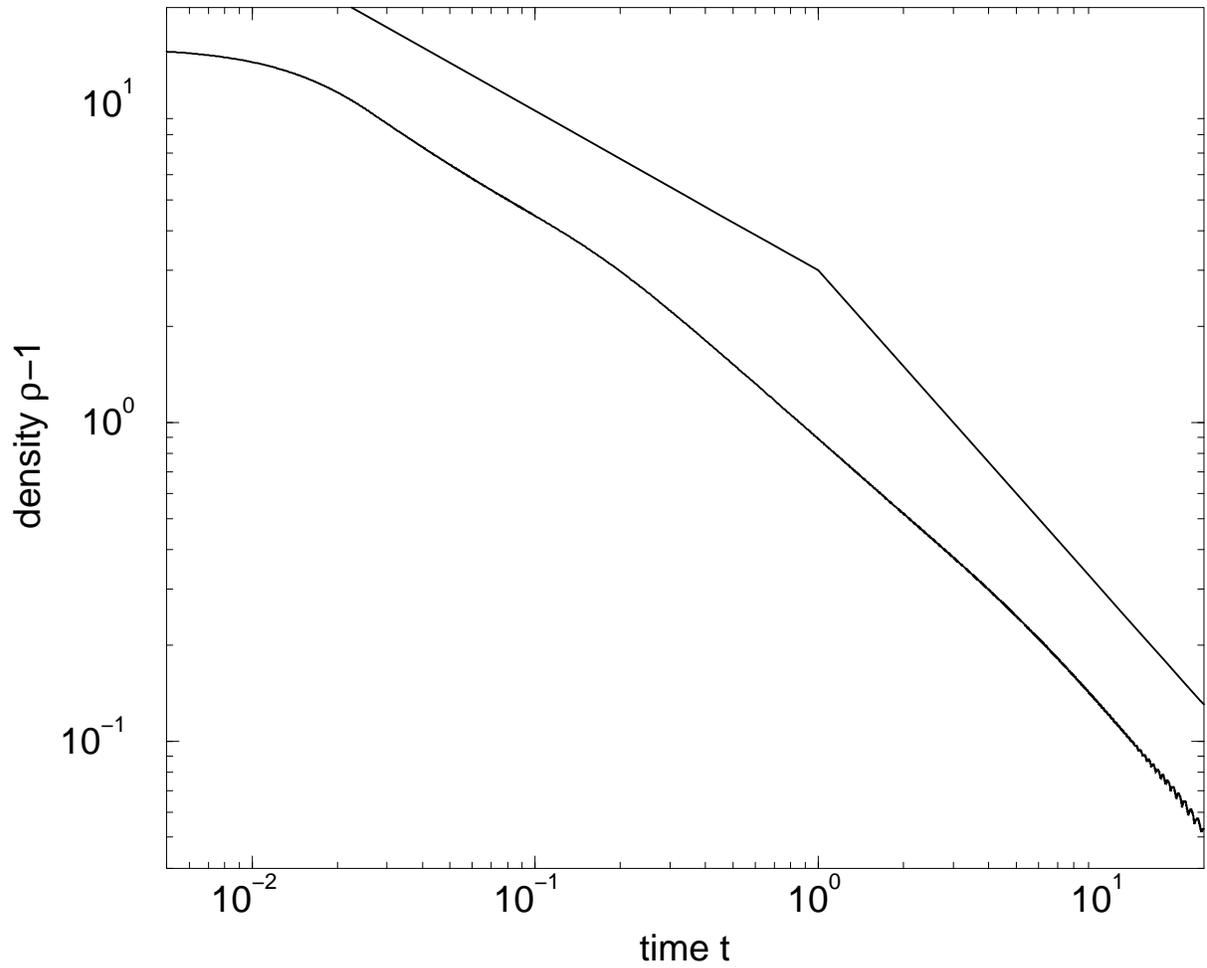} }
\caption{\protect\small The deviation from average density as function of time.
After a short transient, this quantity evolves as a power law in time 
($\rho_p-1 \propto t^{-\mu}$). The two lines above the graphs correspond to
$\mu=0.5$ for small time and $\mu=1$ later.
\label{surden}}
\end{figure}

Figure (\ref{maxi}) and figure (\ref{surden}) show respectively the position 
and the excess density $\rho_p-\rho_0$ of the point (P) as a function of time for the
explosion listed above ($\delta=3$ and $\sigma=1$). 
Both graphs exhibit, after a small transient, two dynamical regimes.
For small time, (such that the density $\rho_p \gg 1$ ), the behavior
is consistent with the following scaling laws:  $R_p \propto \sqrt{t}$
and $\rho_p-\rho_0 \propto t^{-1/2}$ (we define $\nu$ and $\mu$ such
that $ R_p \propto t^\nu$ and $\rho_p-\rho_0 \sim t^{-\mu}$). For different
values of $\delta$ and $\sigma$, $\nu$ varies between $0.46$ and
$0.51$, whereas $\mu$ varies roughly between $0.5$ and $0.6$.  At large time
(in the figures, for $t>1$), we obtain $R_p \propto c_s \cdot t$
 and $ \rho_p-\rho_0 \sim 1/t$ ($\nu=\mu=1$), $c_s$ being the local
 sound velocity.  In this case, as $\delta$ and $\sigma$ change, these
values does not vary significantly (typically between $0.96$ and $1$).

The small time regime can be viewed as an explosive-like regime: the
shock wave is supersonic whereas its density is large ($\rho_p \gg
1$).  For the long time, the system is submitted to two effects: one
for the bubble dynamics, and one for the shock wave that is now
transformed in a spherical divergent sound
 wave (as $\nu=\mu=1$, $\rho_p-\rho_0 \propto 1/R_p$ in fact, as for
 spherical sound wave).  The crossover between these two regimes occurs
when the shock wave looses its supersonic speed (in the simulation
shown here, it happens for $t\sim 1$ unit time). Usually, if a bubble
is nucleated, it occurs also around the crossover time.

\begin{figure}[h]
\centerline{ \epsfxsize=16truecm \epsfbox{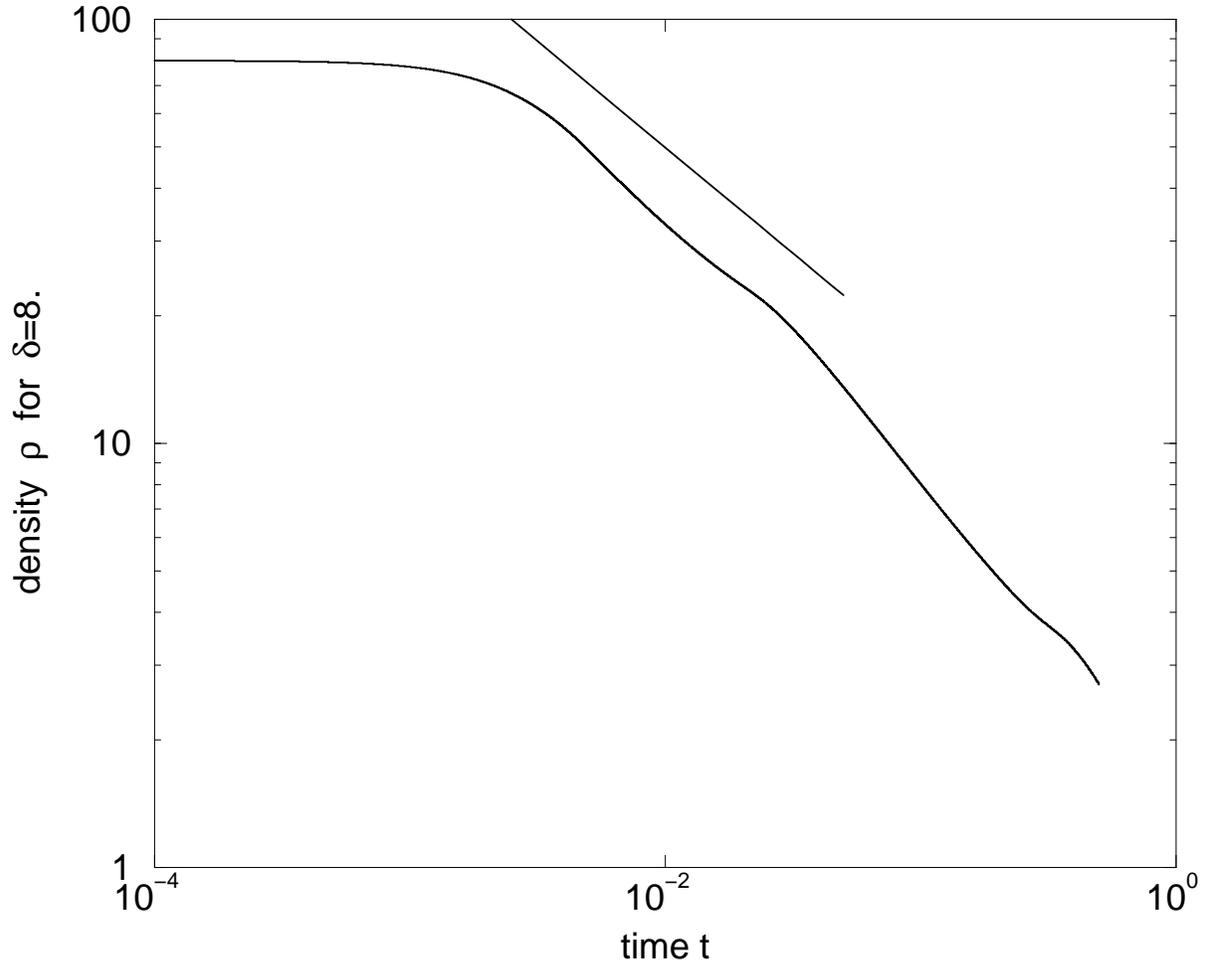} }
\caption{\protect\small The over-density as function of time for a huge explosion
$\delta=8$ and $\sigma=1$. The line indicates the slope $t^{-1/2}$. The small
time behavior coincides with a $\mu=1/2$ regime.
\label{bigexp}}
\end{figure}

In fact, the cross-over between these two regimes is still valid,
 whether a bubble is nucleated backward or not. However, as the
behavior for $R_p$ does not depend on the initial energy (and therefore
on $\delta$ and $\sigma$), the excess density is much more sensible. For
long time, the shock wave is always behaving as a sound wave ($\mu=1$),
although the $\mu=1/2$ property for short time is observed only for big
enough explosion (see figure \ref{bigexp} for $\delta=8$); on the other
hand, for small initial energy, $\mu$ is diverging strongly from
$1/2$.  It actually goes continuously from $\mu=1/2$ for large energies
to $\mu=1$ for small ones (where the shock wave is immediately a sound
wave).

\begin{figure}
\centerline{a) \epsfxsize=7truecm \epsfbox{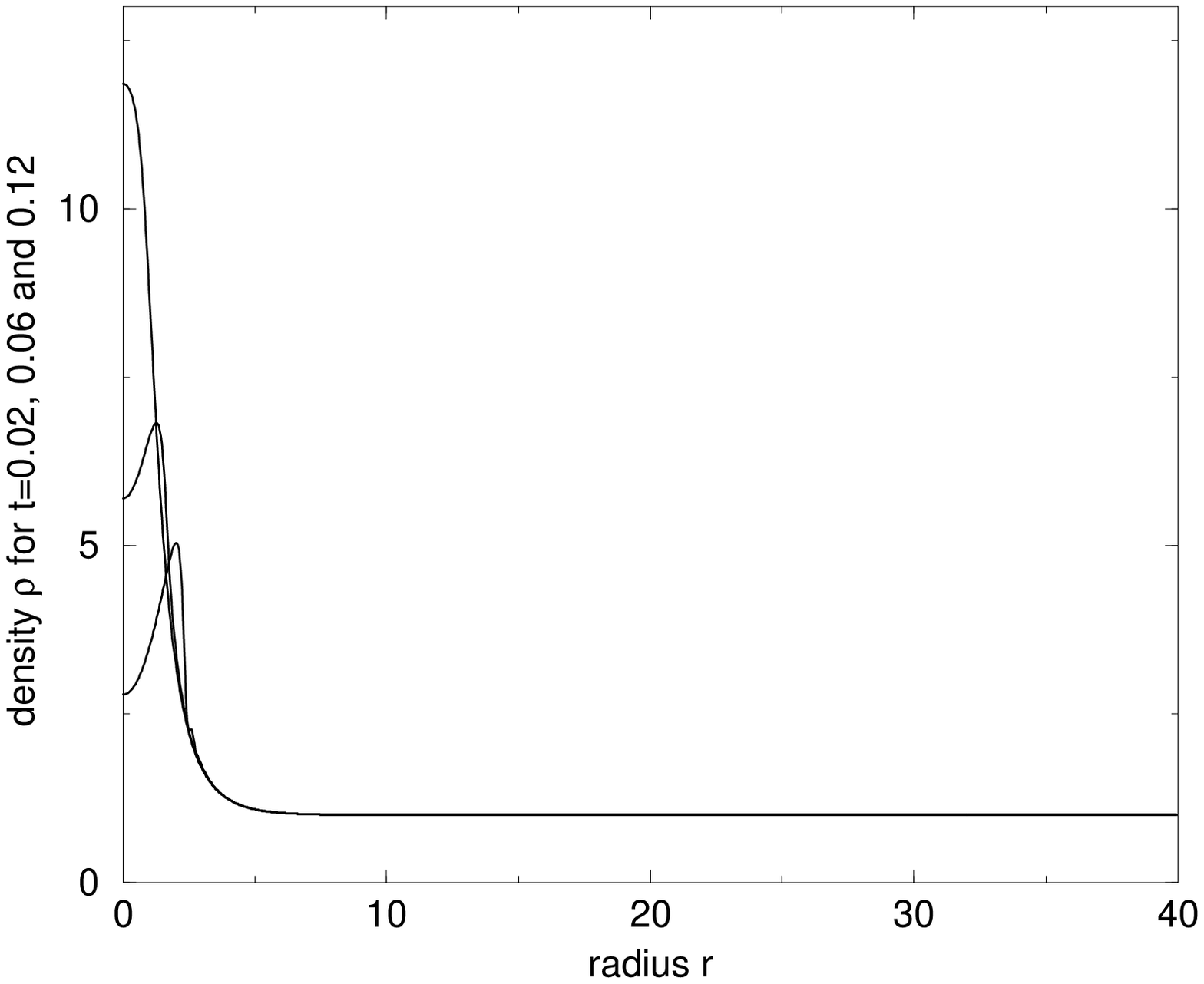} b) \epsfxsize=7truecm
\epsfbox{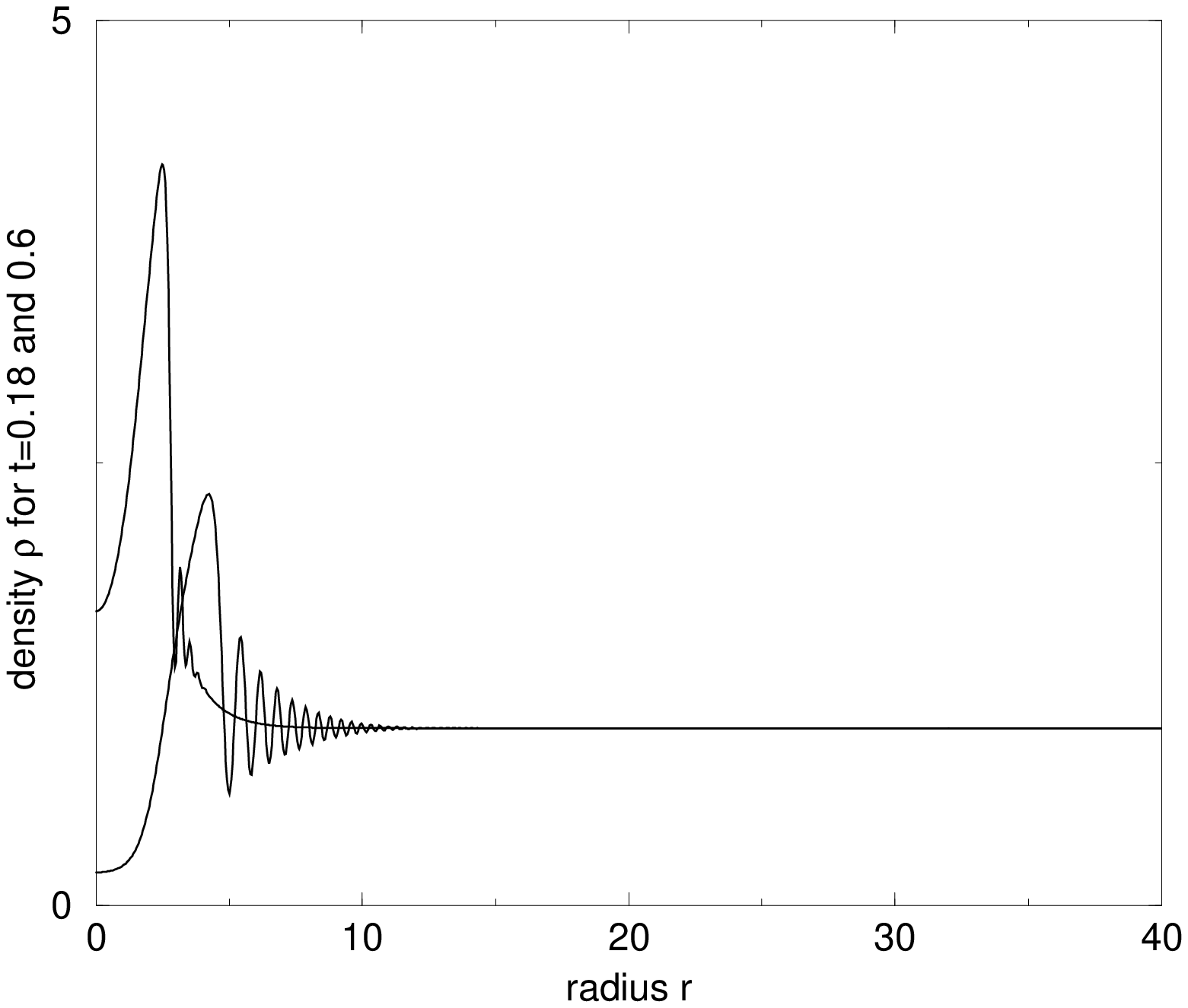} }
\centerline{c) \epsfxsize=7truecm \epsfbox{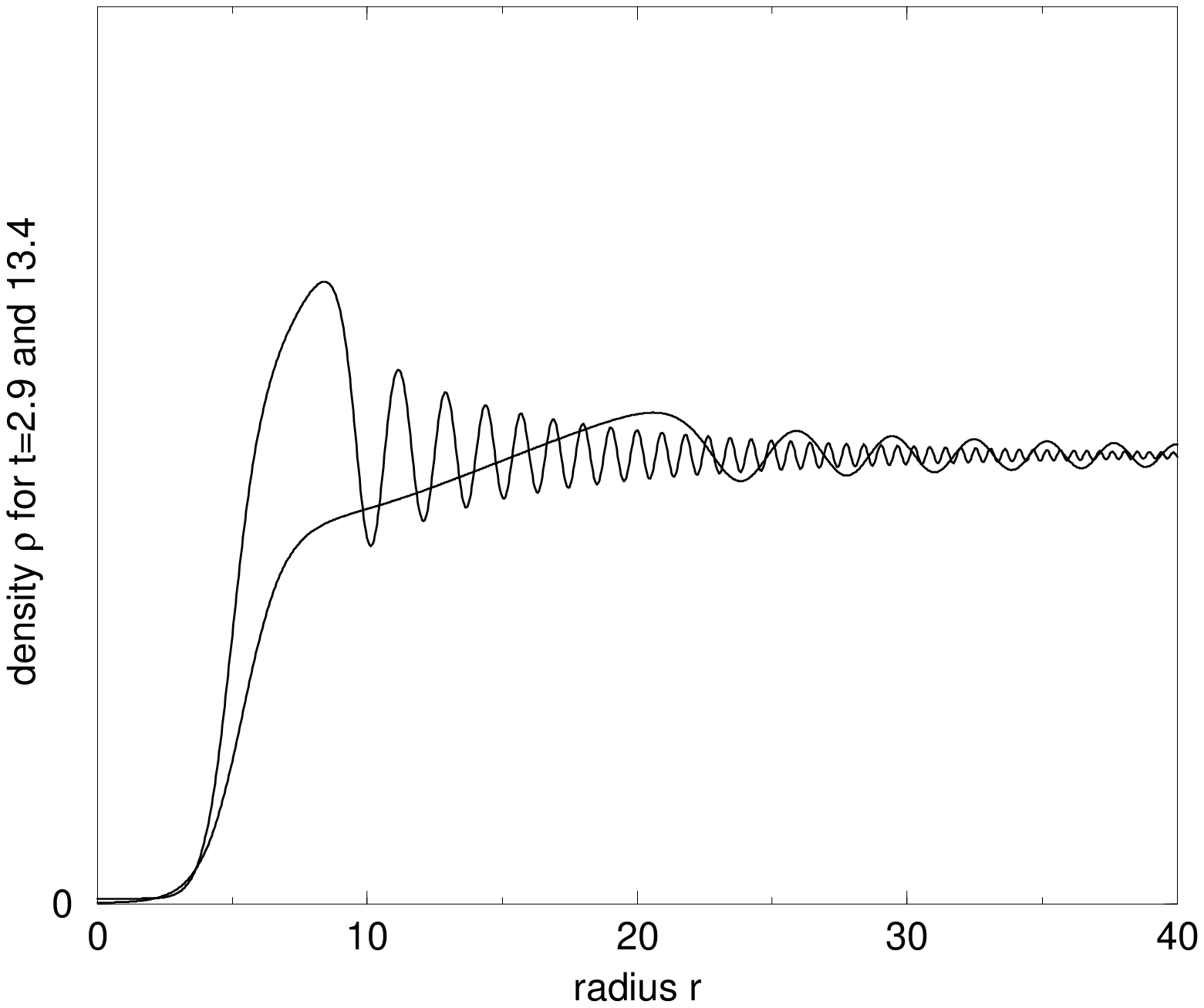} d) \epsfxsize=7truecm
\epsfbox{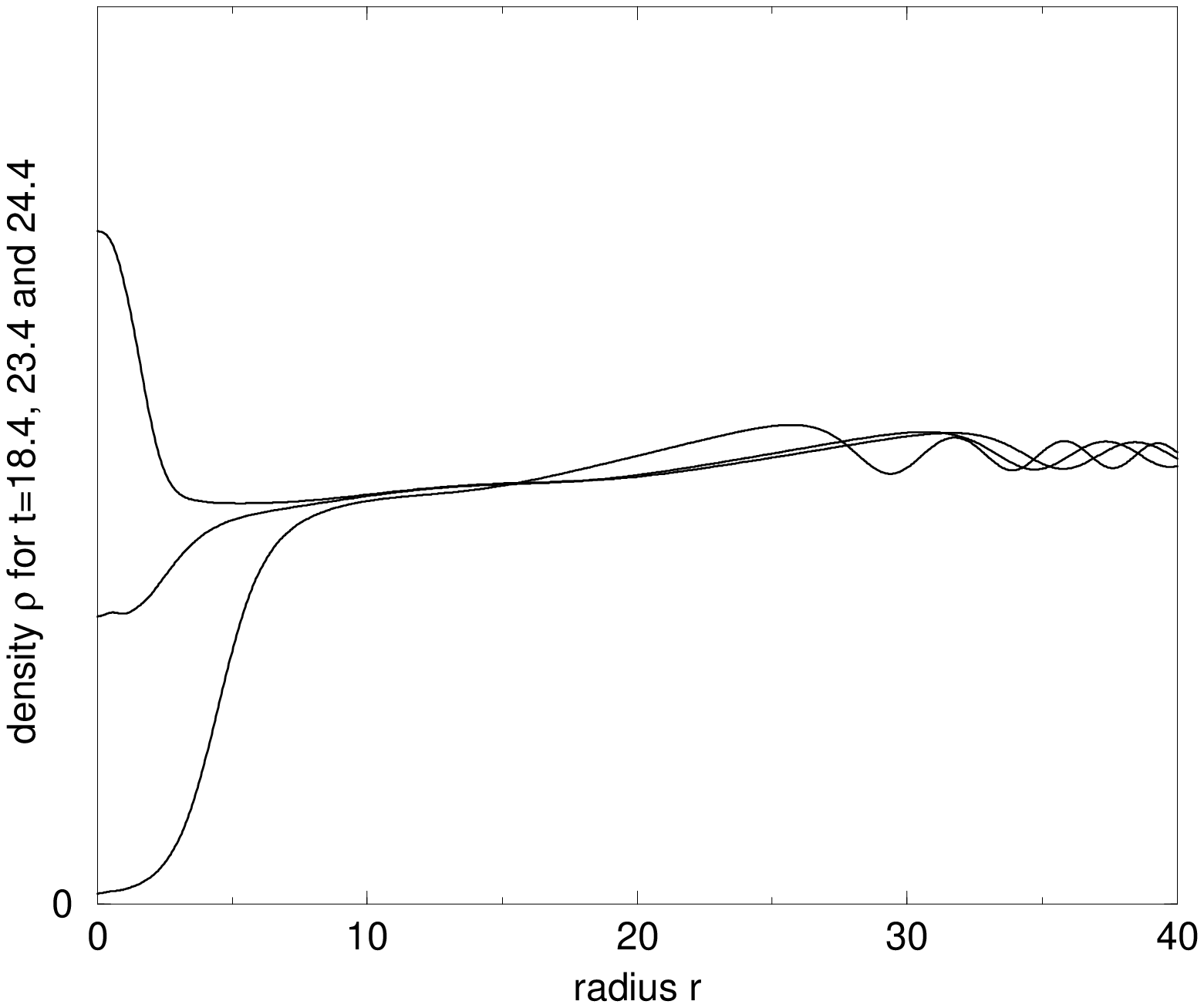} }
\caption{\protect\small Density profile for $\delta=3$ and $\sigma=1$ at 
various time a) for small time, t=0.02, t=0.06 and t=0.12 unit 
time; the explosion gives rise to a shock wave. b) for t=0.18 and 
t=0.6; a bubble appears beside the tensile part of the shock wave; 
oscillating perturbations are emitted by the shock wave. 
 c) for t=2.9, t=7.9 and 13.4; the bubble grows until a maximal radius and 
begins to retract at t=13.4; the shock wave has changed in a spherical sound
wave and is no more moving with the bubble interface.
d) t=18.4, t=23.4 and t=24.4 the bubble collapses, giving rise to a secondary 
explosion. A large part of the energy has been radiated through the spherical
waves.
\label{evolu}}
\end{figure}

As the bubble is nucleated beside the tensile wave, it has its own
dynamics.  One can see clearly on figure \ref{evolu} b) and c) that the
bubble interface moves more slowly that the wave. The wave goes away form
the center, whereas the bubble grows to its maximal radius 
through a Rayleigh-Plesset like dynamics (see \cite{brennen}) and then
collapse under the action of the ambient pressure (figure \ref{evolu}
d)).  We have defined in fact the radius of the gas bubble $R_b$ as the point
where the density reaches a typical low value:
$$ \rho(R_b)= a \cdot \rho_0 $$
We took $a=0.2$ below, although we did check that changing the value of $a$ does not
alter the results.

\begin{figure}[h]
\centerline{ \epsfxsize=16truecm \epsfbox{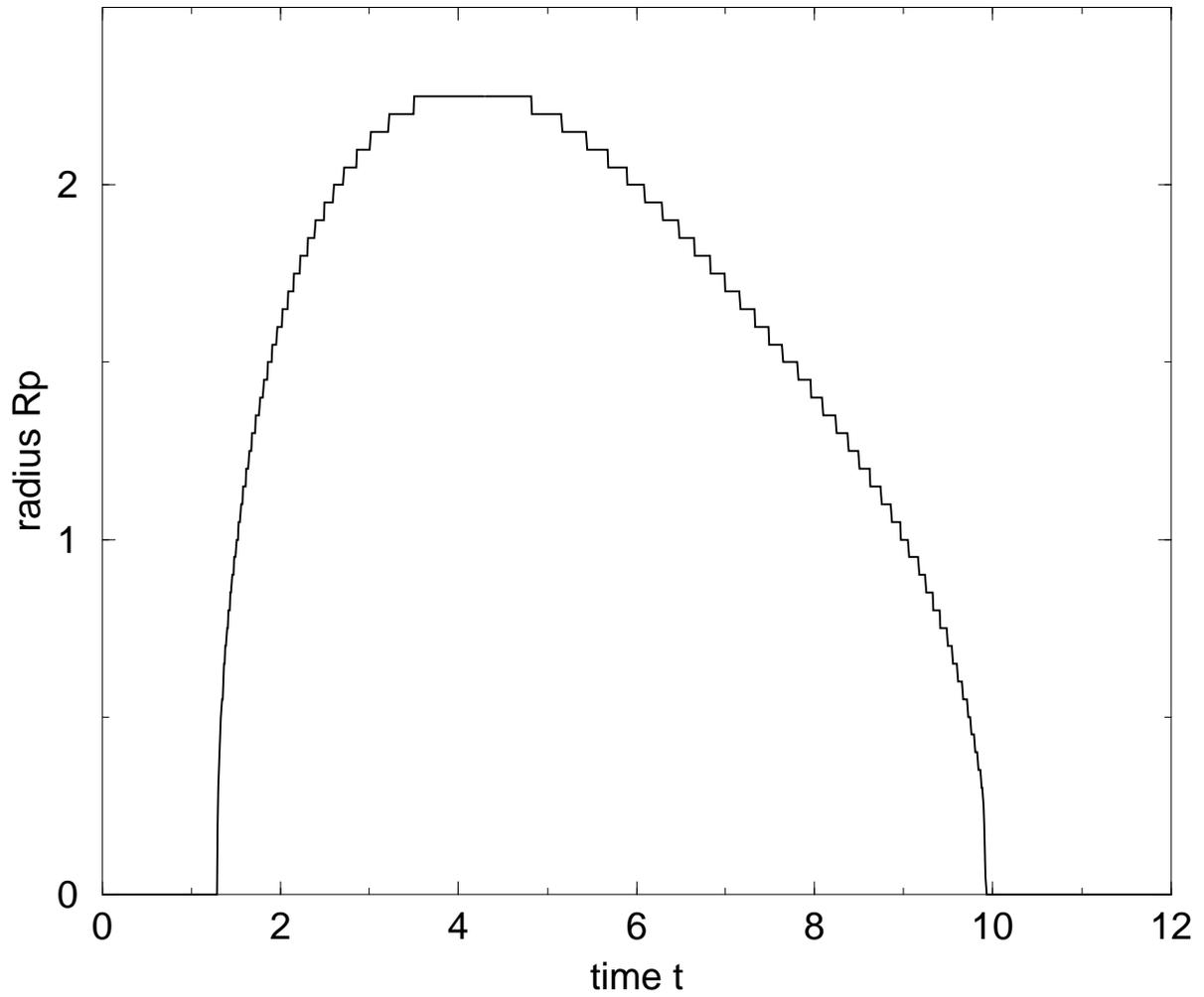} }
\caption{\protect\small Radius of the bubble induced by an explosion as
function of time ($\delta=2$ and $\sigma=1$). 
The growth and the collapse are not symmetric.
Both the growth and the collapse obey the Sedov-Taylor law for the radius 
as function of time.
\label{rayon}}
\end{figure}

Figure (\ref{rayon}) shows the radius of the bubble as
function of time (for $\delta=2$ and $\sigma=1$) and illustrates that:\\
-the bubble is nucleated at time $t^*$. \\
-the growth and the collapse of the bubble are not symmetric. The
collapse stage is slightly longer than that of the growth. We define
$t_c$ as the time when the collapse occurs ($R_b=0$).\\
-the evolution of the radius for short time satisfies for the growth the relation:
$$ R_b(t) \propto (t-t^*)^\frac{2}{5} $$
as well as for the collapse: $R_b(t) \propto (t_c-t)^\frac{2}{5} $.

Last, we computed the ratio between the energy of the secondary peak 
versus the energy of the initial one. We found, depending strongly on the
initial shape of the explosion and also on the initial energy $E_0$, that 
this ratio varies between $0.1$ to $0.25$. Therefore, a large part of the 
energy has actually been emitted by the explosion and only a small part 
transforms in the bubble. In addition, these ratios has been found to be in reasonable 
agreement with the ratios obtained between the rebound bubble and the initial 
one in the experiments \cite{comm}.
 
\section{Analysis of the exponents.}

We will in this section follow the general ideas of explosion
\cite{landau}.  The exponent $2/5$ found for the bubble radius is
famous in explosion and comes naturally from a dimensional analysis.
Actually, the bubble dynamics involves only a small set of variable: the
radius $r$, the time $t$, the energy $E_b$ of the bubble and the
density $\rho_c$. This is an important point: the shape of the bubble
interface is only determined by $\rho_c$ as it is the spherical
solution of SNLS which separates the gas ($\rho=0$) and the liquid
($\rho=\frac{3}{2}\rho_c$) phases. Therefore, only one dimensionless
variable can be obtained from this set:

$$  \frac{Et^2}{\rho_c r^5}$$

consequently we have obtained the well-known dependence of the radius
of the bubble with time:

$$ R_b(t)= \beta \left( \frac{E t^2}{ \rho_c} \right)^{1/5}  $$
where $\beta$ is a constant (determined in fact by the shape of the interface).

\begin{figure}
\centerline{ \epsfxsize=12truecm \epsfbox{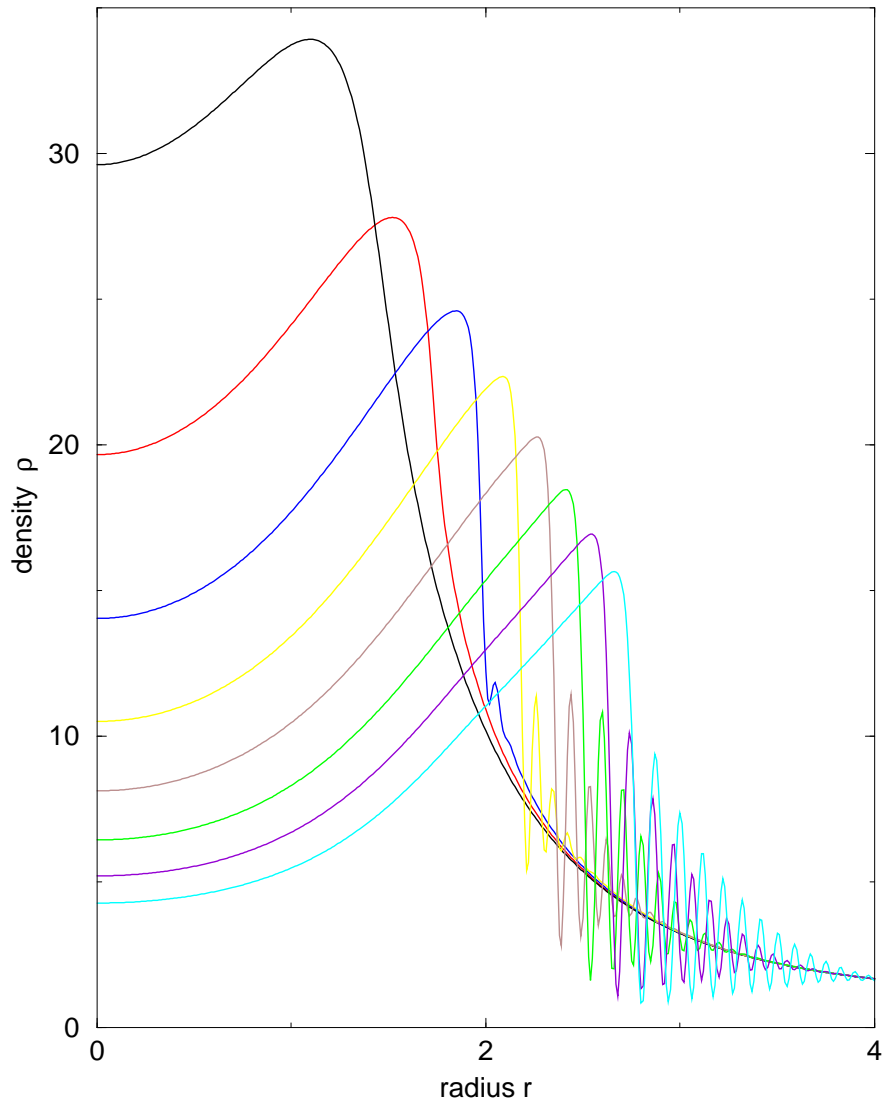}}

\caption{\protect\small density profile of the solution of SNLS for
$\delta=8$ and $\sigma=1$ for different time smaller than $0.05$ unit time.
One can observe a self similar shape in the shock wave as time goes on.
Even if the density obeys a linear profile for intermediate values of
$r$, it is not valid anymore for $r \rightarrow 0$.
\label{rdens}}
\end{figure}

{\it A contrario}, before the bubble is nucleated, the shock wave obeys
different scaling and such a dimensionless variable is not unique. Indeed, 
the density of the shock waves also varies now. But, as figure (\ref{rdens}
shows, for huge explosion (here $\delta=8$), a self-similar shape in the 
density profiles can be observed. We will therefore limit
our investigation to big
explosions ($\delta \gg1$) and to time such that $\rho \gg 1$ everywhere
inside the shock region; in addition, the following remarks are needed:

 -first of all, a self-similar solution cannot conserve both energy and mass.
 (this follows from the fact that mass is an integral of $|\psi|^2$, whereas
energy is an integral of $|\psi|^6$).

 -consequently, the limit of infinitely thin explosion is either finite
 energy with zero mass or finite mass and infinite energy.

 - due to the condition at infinity,
($\rho=\rho_0$), a flux of mass is continuously entering the
self-similar region as the time goes on, whereas such mass density has
zero energy density.

Balancing terms in Bernoulli equation (\ref{berns}) motivated us to 
look at solutions such that (the cubic term is neglected versus the 
quintic one in SNLS):

\begin{equation}
\rho(r,t)=\frac{1}{\sqrt{t}}h^2(\xi) \quad \phi(r,t)
=g(\xi) \quad {\rm with} \quad \xi=\frac{r}{\sqrt{t}}.
\label{simil1}
\end{equation}

In this approach, $\xi$ is the self similar variable. It has the same 
structure as the one elaborated for the heat equation. In both case, it
comes from the balance between the time derivative and the laplacian term.
Therefore, the position of the peak of density is defined by $\xi=a$, 
where $a$ is determined by the initial condition and is time independent.
Equation (\ref{simil1}) is valid everywhere except that because of the
assumption ($\rho \gg 1$), it is relevant and consistent only near the
shock wave region. Also, in our regime of interest, we have $a \gg 1$.

It follows from this analysis that:
$$ R_p(t)=a \sqrt{t} \quad {\rm and} \quad \rho_p-\rho_0=\frac{h^2(a)}{\sqrt{t}}
$$

which are in good agreement with the numerics. Notice that if the
quintic term were negligible versus the cubic one in SNLS, the scaling
for the density $\rho_p-\rho_0$ would have been in $1/t$. It explains
why when the energy of the explosion is lowered, the value of $\nu$
goes continuously from $1/2$ to $1$.

Finally, with these scaling laws, the mass inside the shock sphere
evolves linearly in time, whereas the available flow of particles goes
like the square root of time. Then with respect to mass conservation,
these scaling laws cannot apply as time goes to zero, when the entering
flow of particles cannot balance the growth of mass of the solution.
This gives a critical time under which the self-similar solution is not
valid. Numerically, it is difficult to see if the transient that we
observe initially is due to this constraint or to numerical ones.

The radial velocity $v$ is obtained through:
$$ v(r,t)= \frac{\xi}{r}g'(\xi) $$

and the Bernoulli equation (\ref{berns}) reads:

\begin{equation}
\frac{\xi}{2}g'(\xi)=-\frac{1}{2h(\xi)} \left( h''(\xi)+2
\frac{h'(\xi)}{\xi} \right) + \frac{1}{2} (g'(\xi))^2+h^4(\xi)
\label{selb}
\end{equation}

Another equation is needed for describing the dynamics; the natural one
would be the mass conservation. In fact, the integral equation for the
conservation of energy in a sphere of constant radius in the self
similar variable ($\xi={\rm constant}$) is more useful:

\begin{equation}
v\cdot \left( (\nabla\sqrt{\rho})^2+\frac{1}{2}\rho v^2-\frac{1}{2}\sqrt{\rho}
\Delta\sqrt{\rho}+\rho^3 \right)+\frac{1}{2}\sqrt{\rho}\nabla(\sqrt{\rho})\Delta
\phi=\frac{\xi}{a} R'(t)\epsilon(\rho,v)
\label{selfe}
\end{equation}

$\epsilon $ being the density of energy:
$$ \epsilon(\rho,v)=\frac{1}{2}\rho v^2+\frac{1}{2}(\nabla\sqrt{\rho})^2+
\frac{\rho^3}{3} $$
Equation (\ref{selfe}) reflects the fact that the increase of energy inside the 
sphere of radius $\xi$ is balanced by the enthalpy flux.

An exact solution of the system can be found by neglecting the quantum 
pressure: 

\begin{eqnarray}
 v=\frac{\xi}{4\sqrt{t}} \quad {\rm and} \quad \rho=
\sqrt{\frac{3}{2}}\times \frac{\xi}{4\sqrt{t}} \nonumber \\
{\rm corresponding \quad to} \quad g(\xi)=\frac{(\xi^2-a^2)}{8} \quad {\rm 
and} \quad h^2(\xi)=\sqrt{\frac{3}{2}}\times \frac{\xi}{4}
\label{self1}
\end{eqnarray}

These solutions also satisfy exactly the continuity equation (\ref{conts}).
Identifying the energy of such a solution with the initial energy $E_0$ gives:

$$ E_0 = \frac{16}{15} \delta^6 \sigma^3 =\frac{1}{64}\sqrt{\frac{3}{2}} a^6$$
and therefore: 
$$ a \propto \sqrt{\sigma} \delta $$

\begin{figure}
\centerline{a) \epsfxsize=8truecm  \epsfbox{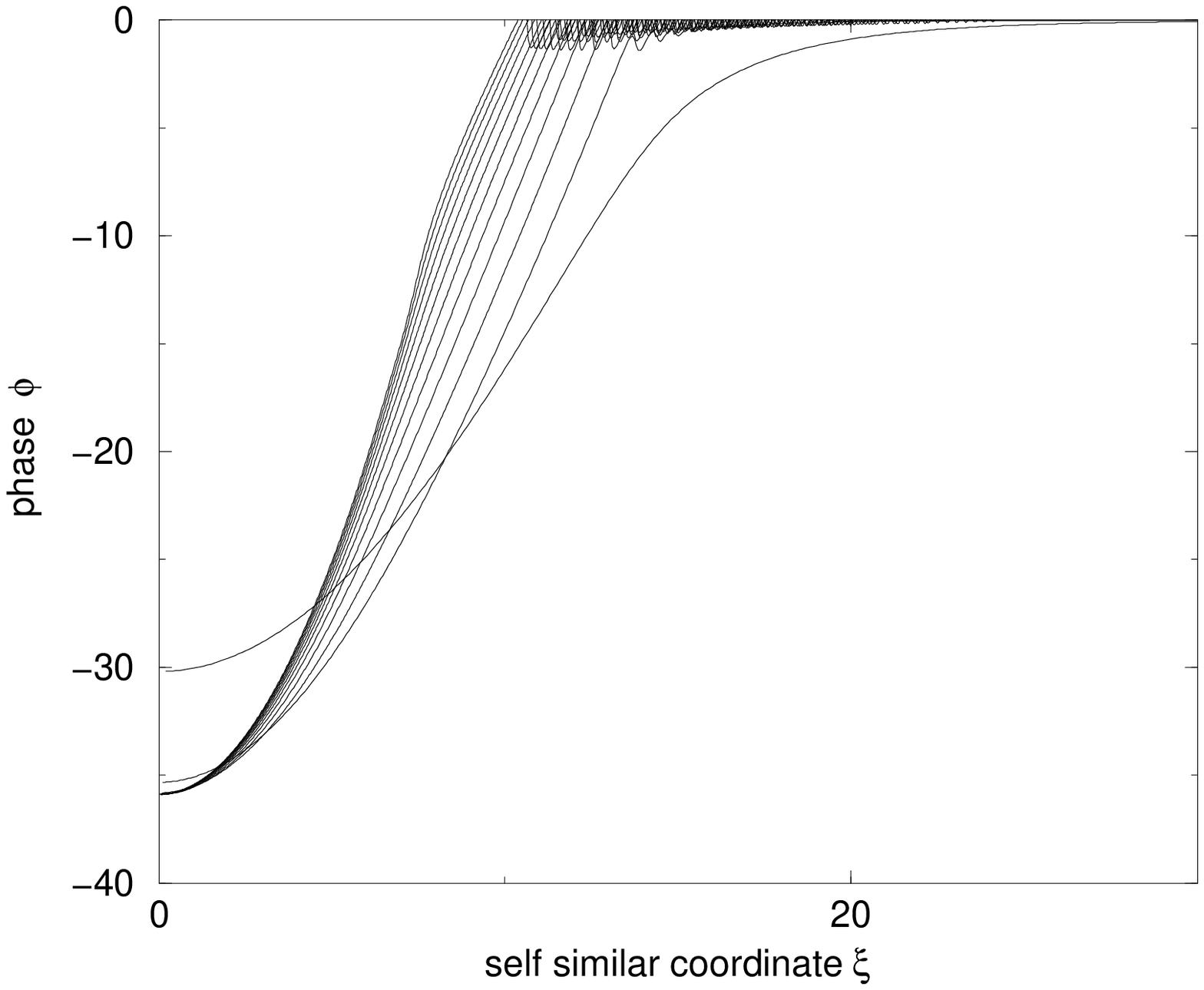} b) 
\epsfxsize=8truecm \epsfbox{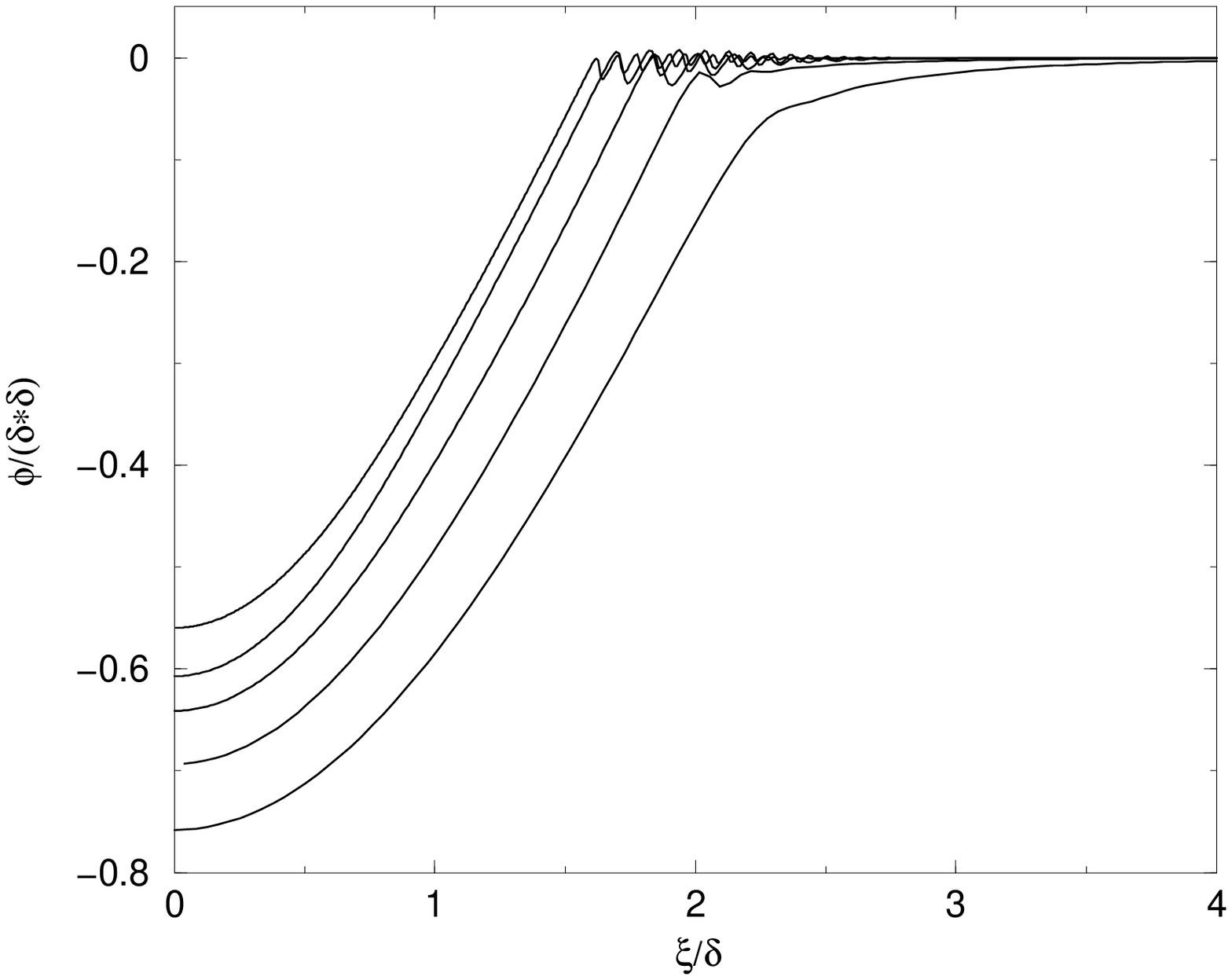}}
\caption{\protect\small a) phase $\phi$ as a function of $\xi$ the self 
similar variable, for different time (from $t=0.01$ to $t=0.2$; $\delta=8$).
Except the two first curves, which correspond to the lowest time, where the 
dynamics is still in the transient, the different curves almost coincide 
for $\xi \sim 0$ and have a parabolic shape, as predicted by the simple
self similar approach. b) for different $\delta$, $\phi/\delta^2$ as function 
of $\xi/\delta$; according to the self similar analysis, these curves should 
coincide.
\label{fpha}}
\end{figure}

Figure (\ref{fpha} a) shows the phase as function of $\xi$ at
different times for $\delta=8$ and $\sigma=1$; it offers a good 
agreement with the parabolic analytical solution of equation 
(\ref{self1}).
Particularly, the minimum of the curve (at $\xi=0$) does not change its
value as time goes on, as predicted by the self similar solution. In
addition, figure (\ref{fpha} b)) shows for different $\delta$ (with
$\sigma=1$), $\phi(\xi/a)/a^2$; the solution (\ref{self1}) suggests that 
these curves should coincide. The numerical results are less accurate than 
the one for a given $\delta$, although they are still reasonable.

On the other hand, the density profile provided by the solution (\ref{self1}) 
does not agree well with 
the numerics (see figure \ref{rdens} for $\delta=8$): although the linear 
behavior for $\rho$ is acceptable in an intermediate range ($1\ll\xi \ll a$),
 there are strong differences
for $\xi \sim 0$ and $\xi \sim a$. For $\xi <1$, the quantum 
pressure becomes dominant in the self similar set of equations (\ref{selb}) 
 and (\ref{selfe}). For
$\xi \sim a $ the solution is in fact maximal and also has to match
 with the condition at infinity.

As the numerical simulations show a good agreement with the self similar
solution for the phase, we will assume the solution (\ref{self1}) to be
exact for $\phi$ and therefore we will
 try to find a more accurate solution for $\rho$, particularly in
restoring the quantum pressure. After rescaling $h$ and $\xi$ for
convenience ($h(\xi)=\frac{3^{1/8}}{8^{1/4}} f(3^{1/4}\xi/2)$), the
problem reads:

\begin{equation}
f''(u)+2\frac{f'(u)}{u} =f^5(u)-u^2 f(u)
\label{self2}
\end{equation}

The boundary of the self-similar region being then defined by
$u=a'=3^{1/4}a/2$.  Even if the equation is relevant only for $u<a'$, it is
interesting to study it for all values of $u$.
 Using Mathematica\cite{wolf}, it is straightforward to analyze
the class of numerical solutions of (\ref{self2}) with the physical
initial conditions $f(0)=A$ and $f'(0)=0$. Depending on the value of
$A$ compared to a critical value $A_c =0.98473...$, we obtain the following
behaviors:

-for $A<A_c$ the solution oscillates and goes to zero at infinity.

-for $A>A_c$ the solution includes a singularity in the real plane.

\begin{figure}
\centerline{ \epsfxsize=16truecm \epsfbox{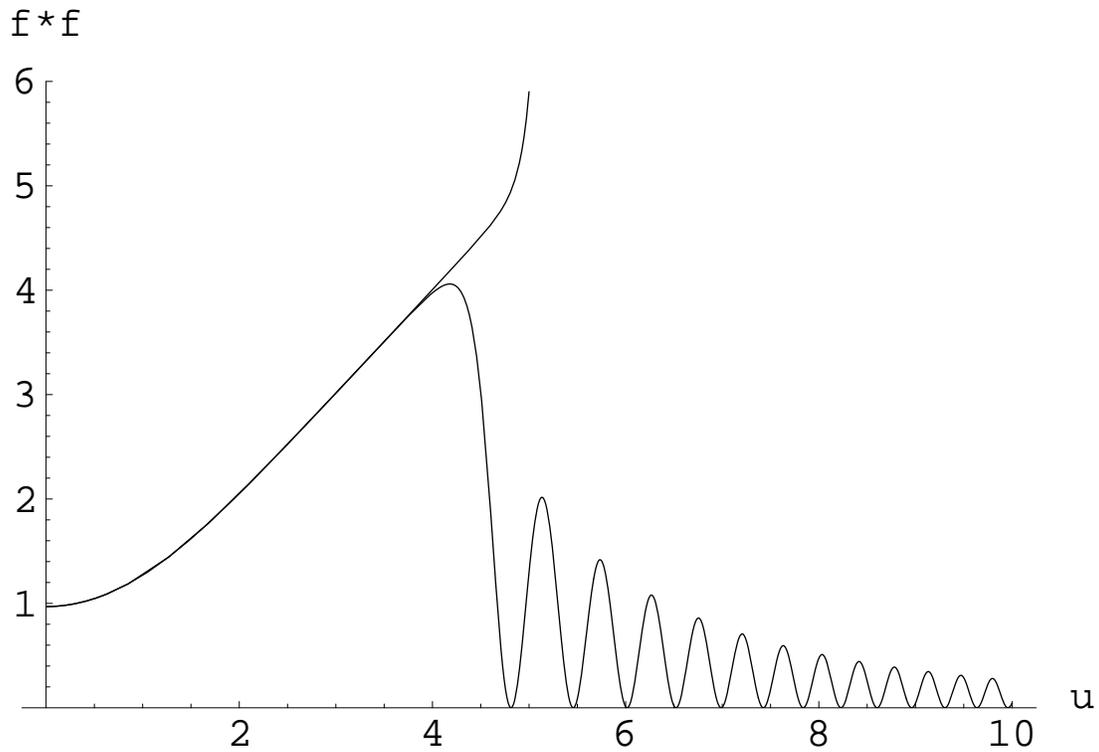}}
\caption{\protect\small Numerical solution of equation \ref{self2} using
the shooting method, where $f'(0)=0$ and $f(0)=A$. For the lower curve,
$A=0.98473178<A_c$ and the upper one is for $A=0.98473178435>A_c$.
\label{matself}}
\end{figure}

Figure (\ref{matself}) shows the solution $f(u)^2$ as a function of $u$ for
two values of $A$: one slightly smaller than $A_c$, the other slightly larger. 
$A_c$ has been evaluated through the shooting method. The solution for $A=A_c$ 
will behave linearly in $u$ as $u \rightarrow \infty$, in agreement with the 
solution (\ref{self1}).

The solutions for $A<A_c$ appear to be in good agreement with the numerical solution 
of SNLS (see again figure \ref{rdens}), and we will therefore assume that it
describes the self similar solution. The exact value of $A$ will be a function 
of $a$ and consequently of $E_0$. In addition, the matching between this solution
around $u \sim a'$ with the sound wave solution of SNLS (at density $\rho_0$) 
inform us about the wave number of the oscillations emitted by the solution.
The oscillating solution at large $u$ for $A<A_c$ can be obtained easily by 
neglecting $f^5$ when $u \rightarrow 
\infty$; then the equation:
$$ f''(u)+2\frac{f'(u)}{u} +u^2 f(u)=0 $$
has an exact solution in terms of the Bessel functions of the first kind 
$J_n(\cdot)$. It reads:

$$f(u)=\frac{C_1 \cdot J_{-1/4}(u^2/2)+C_2 \cdot J_{1/4}(u^2/2)}{\sqrt{u}} $$
$C_1$ and $C_2$ being two constants.

For large $x$, the behavior of $J_{\pm 1/4}(x)$ is simple:
$$ J_{\pm 1/4}(x) \sim \frac{{\rm cos}(x+\varphi)}{\sqrt{x}} $$
where $\varphi$ is a constant.\\
Therefore we obtain the following behavior for $f(u)$ for large $u$ 
(more precisely for $u \gg 1$):

$$ f(u) \propto \frac{cos(\frac{u^2}{2}+\varphi)}{u^{3/2}} $$

So that, in the neighborhood $u \sim a'$, we have:
$$ f(a'+\delta x) \propto \frac{cos(a' \cdot \delta x + \varphi')}{(a')^{3/2}}$$

This corresponds to oscillation for the density $\rho$,
 around $r=a\sqrt{t}$, with wave number $k(t)$:

 $$k(t)=\frac{\sqrt{3}a}{\sqrt{t}}$$

This simple matching between the self similar solution and the density at 
infinity through excitation waves gives in fact the wave number of these 
radiations as function of time. 

\section{Renormalisation Group calculations.}

Through renormalisation group (RG) calculations\cite{golden}, one can
describe precisely the behavior of $f(u)$ at both edges of the shock wave $u=0$
and $u=a'$.

{\bf for $\xi \rightarrow 0$:} introducing an arbitrary small parameter $\epsilon$,
through $ \eta = \frac{u}{\epsilon}$, the following equation for $m(\eta)=
f(u)$ is obtained:
\begin{equation}
m''(\eta)+2\frac{m'(\eta)}{\eta}=\epsilon^2 m^5(\eta)-\epsilon^4 \eta^2m(\eta)
\label{eqren1}
\end{equation}
Then $m(\eta)$ is expanded as a series in $\epsilon^2$:
$$ m(\eta)=m_0(\eta)+\epsilon^2m_1(\eta)+\epsilon^4m_2(\eta)+... $$

The system of equation for $m_0$ and $m_1$ reads:

\begin{eqnarray}
m_0''+2\frac{m_0'}{\eta}=0 \nonumber \\
m_1''+2\frac{m_1'}{\eta}={m_0}^5 \nonumber \\
\nonumber
\end{eqnarray}

The boundary conditions for $m$ come from the constraint $m'(0)=0$ (no 
flux at $\eta=0$) and from the asymptotic behavior at large $\eta$, which 
will act as a matching condition for the remaining constant of integration.

Using $m'(0)=0$, one obtains:
$$ m(\eta)=A(0)+\frac{\epsilon^2\eta^2}{6} A(0)^5 $$
$A(0)$ being the constant of integration at $t=0$\footnote{formally, the other solution 
of this second order differential equation $m=B/\eta$ should be kept for the 
formal solution of the problem because the boundary condition $m'(0)=0$ has to
be imposed only for the final results; however, as we checked that this term has
zero contribution at the order of the analysis, it has been forgotten for 
convenience.}.

This formula loses its validity for $\epsilon \eta > 1$ because the 
perturbative expansion of $m$ is meaningless. ($\epsilon^2m_1$ is no longer a 
small correction to $m_0$ in this regime).

The renormalisation group (RG)  method consists in considering the integration 
constant as being slightly dependent on $\eta$ through the change:
$$ A(0)=Z(0,\mu)A(mu) \quad {\rm with} \quad Z(0,\mu)=1 +\sum a_n(\mu)
\epsilon^{2n} $$
$Z$ is called  the multiplicative renormalization constant and $a_n$ is
chosen such that the solution $m(\eta)$ has the same structure for all 
$\mu$:
$$ m(\eta)=A(\mu)+\frac{\epsilon^2 (\eta^2-\mu^2)}{6} A^5(\mu) $$
This implies that:
 $$a_1=-\frac{\mu^2A^4(\mu)}{6} $$
Obviously $m(\eta)$ should not depend on $\mu$. Then, the so-called RG equation
 $\partial_\mu m(\eta)=0$ gives:
$$ \frac{dA}{d\mu}=\frac{1}{3}\epsilon^2 \mu A^5 $$
whose solutions read:
$$A(\mu)=\left(\frac{6}{\chi_0^2-\epsilon^2\mu^2} \right)^{1/4} $$

Substituting $A(\mu)$ into the expression for $m(\eta)$ and putting $\mu=\eta$ 
gives the following formula for the initial function $f$:
$$ f(u)=\left(\frac{6}{\chi_0^2-u^2} \right)^{1/4} $$
$\chi_0$ is a constant of integration to be determined. Notably, we have obtained 
the kind of divergence that occurs for $A>A_c$.

{\bf For $\xi \rightarrow a$:} the same treatment works in the limit $a'
\rightarrow \infty$ by writing:
$$ f(u)= \sqrt{a'}+\frac{m(a'\cdot(u-a'))}{(a')^2} $$
($1/a'$ has the same feature that $\epsilon$ had formerly and the $1/(a')^2$ 
factor has been introduced in order to neglect the non linear terms for the
first order correction).

Eventually, it gives the following behavior ($A_0$ being a constant):

$$f(u)=\sqrt{a'}-\frac{A_0}{(a')^2} e^{\frac{9(u-a')}{8a'}} 
{\rm cosh}(a'(u-a')-\frac{(u-a')^2}{4})$$

$A_0$ and $\chi_0$ are yet to be determined. This is usually accomplished through a
matching condition written in an intermediate region between these two 
regimes. In our special problem, an easy way to perform this would be to match 
both solution in the region $ 1 \ll u \ll a'$ where we know that the solution 
reads $f(u) \sim \sqrt{u}$.
In fact, the first order RG calculations is not accurate enough to realize 
such a matching; but the goal of these calculations was actually to have a reasonable idea of the
shape of the solution near both edges. The whole matching process would 
need a more complete analysis of the RG theory.

\section{Conclusion-Acknowledgments}

Finally, as the shock emits waves, the energy of the self similar
solution cannot be constant in time. Also, the cubic term has been
neglected for finding the above self similar solutions. This should
really be incorporated into a pertubative analysis by considering $a$
(or equivalently $E_0$) as being slowly dependent on the time. Then,
through a solubility equation, an evolution equation should be found
for $a$ \cite{dyach}. This detailed analysis, which would provide for
example the amplitude of the emitted waves, will be the subject of a
further work.

It is a pleasure for me to thank Yves Pomeau, Leo Kadanoff, Shankar 
Venkataramani and Alberto Verga for helpful discussions and for their interest
on this paper. This work has been supported in part by
the ONR grant: N00014-96-1-0127 and the MRSEC with the National Science 
Foundation DMR grant: 9400379.

\newpage

\newpage 

\newpage 

\newpage 

\newpage 

\newpage 

\newpage 

\newpage 

\newpage 

\end{document}